\documentclass{PoS}
\usepackage{subfigure} 
\usepackage{amsfonts} 
\usepackage{amssymb} 
\usepackage{amsmath} %

\title{PDFs in small boxes}

\ShortTitle{PDFs in small boxes}

\author{Ra\'ul A. Brice\~no\\
       Old Dominion U. and Jefferson Lab, USA\\
       Department of Physics, Old Dominion University, Norfolk, Virginia 23529, USA\\
       E-mail: \email{rbriceno@jlab.org}}

\author{\speaker{Juan V. Guerrero}\\
	Hampton University, Hampton, Virginia 23668, USA\\
	Thomas Jefferson National Accelerator Facility, 12000 Jefferson Avenue, Newport News, Virginia 23606, USA\\
	E-mail: \email{juanvg@jlab.org}}    

\author{Maxwell T. Hansen\\
	Theoretical Physics Department, CERN, 1211 Geneva 23, Switzerland\\
	E-mail: \email{maxwell.hansen@cern.ch}}

\author{Christopher J.~Monahan\\
	Institute for Nuclear Theory, University of Washington, Seattle, Washington 98195, USA\\
	E-mail: \email{cjm373@uw.edu} \thanks{Preprint INT-PUB-18-053}}

\abstract{PDFs can be studied directly using lattice QCD by evaluating  matrix elements of non-local operators.  A number of groups are pursuing numerical calculations and investigating possible systematic uncertainties. One systematic that has received less attention is the effect of calculating in a finite spacetime volume. Here we present first attempts to assess the role of the finite volume for spatially non-local operators. We find that these matrix elements may suffer from large finite-volume artifacts and more careful investigation is needed.}

\FullConference{The 36th Annual International Symposium on Lattice Field Theory - LATTICE2018\\
		22-28 July, 2018\\
		Michigan State University, East Lansing, Michigan, USA.}

\begin{document}

\section{Introduction}

Understanding the internal structure of nucleons remains an important challenge in hadronic physics, despite the significant advances made in this area over the last years. The recent 12 GeV upgrade at JLab and a future electron-ion-collider will make it possible to access the inner structure of hadrons experimentally, in more detail than ever before. However, the interpretation of the resulting data will be more productive if it is accompanied by a complete and detailed theoretical understanding. Within quantum chromodynamics (QCD), information on the internal structure of hadrons is mainly encoded in parton momentum distributions of the hadron constituents, the quarks and gluons, such as transverse momentum dependent distributions and collinear parton distribution functions (PDFs).

Lattice QCD has the potential to provide first principles, non-perturbative, and fully-systematic predictions of the $x$-dependence of PDFs that will shed new light into hadron structure. 
In the last decade,
several ideas has been proposed to achieve this. 
Some of these require the evaluation of matrix elements of non-local operators, for example two quark fields connected by a Wilson line \cite{Ji:2013dva,Radyushkin:2016hsy,Radyushkin:2017cyf}
or two currents separated in space \cite{Braun:2007wv,Bali:2017gfr,Ma:2017pxb} (for a full review, see, for example, \cite{Monahan:2018lon}).

There has been an intense effort dedicated to investigating all possible systematic uncertainties that arise in these kinds of calculations, with the possible exception of finite-volume effects. 
We can interpret the finite volume as a modification to the infrared scales of the theory. Thus finite-volume effects are naturally determined using hadrons as degrees of freedom since quarks cannot propagate long distances. This is nicely illustrated with a simple example: a nucleon, $N$, in a finite box of size $L$. If we impose periodic boundary conditions, the finite-volume artifacts can be understood to arise from the interactions of the hadron with its mirror images. For instance, if we approximate the long distance part of the two-nucleon potential by a Yukawa potential with just $\pi$ exchanges, the finite-volume correction to the mass of the nucleon can be estimated by 
  \begin{equation}
m_N(L)- m_N(\infty) \sim   \int d^3 \textbf x \, \psi_N(\textbf x)  V(\textbf x)  \psi_N(\textbf x + L \textbf e)   \sim e^{-m_\pi L} \,,
\end{equation} 
%
where $ \psi_N(\textbf x)$ is the nucleon wavefunction. This is sufficient to find the now-standard result that scale $m_\pi L$ encodes the  finite-volume artifacts in the energies of stable states~\cite{Luscher:1985dn}.

For the case of a matrix element of two currents separated in space, we {\em a priori} expect substantial finite-volume artifacts. In a finite and periodic volume, the matrix elements would have to be periodic with respect to the separation, $\xi$, of the currents. This is in contrast to the infinite-volume expectation that this matrix element would decay with respect to the separation. 
In this case, there are two IR length scales, the box size $L$ and the nonlocality of the operator $\xi$, so we expect finite-volume artifacts that depend on
both, and it is important to understand in which combination these enter.

In this work, we study and quantify the finite-volume artifacts associated with these matrix elements in the framework of a toy theory. The full details of this analysis are given in Ref.~\cite{Briceno:2018lfj}.

\section{Toy model and set up}

In order to quantify the finite-volume corrections to QCD observables computed in lattice QCD, we should work in the framework of a low-energy effective field theory (EFT), with hadrons as degrees of freedom. Here we consider a scalar field theory, inspired by chiral perturbation theory ($\chi$PT), intended to capture the basic features of the finite-volume artifacts. 

\begin{figure}[t]
	\centering
	\subfigure[]{\includegraphics[width =1\textwidth]{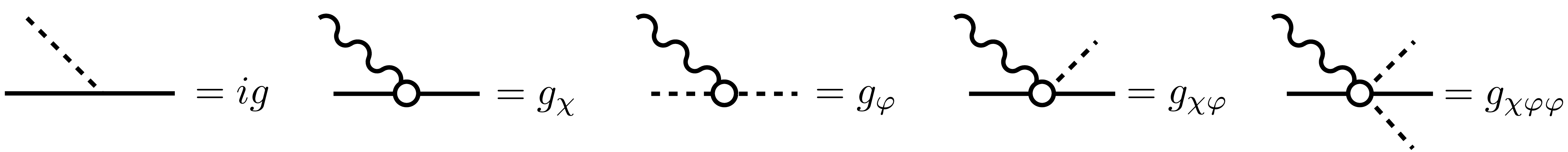}}
	\subfigure[]{\includegraphics[width =.3\textwidth]{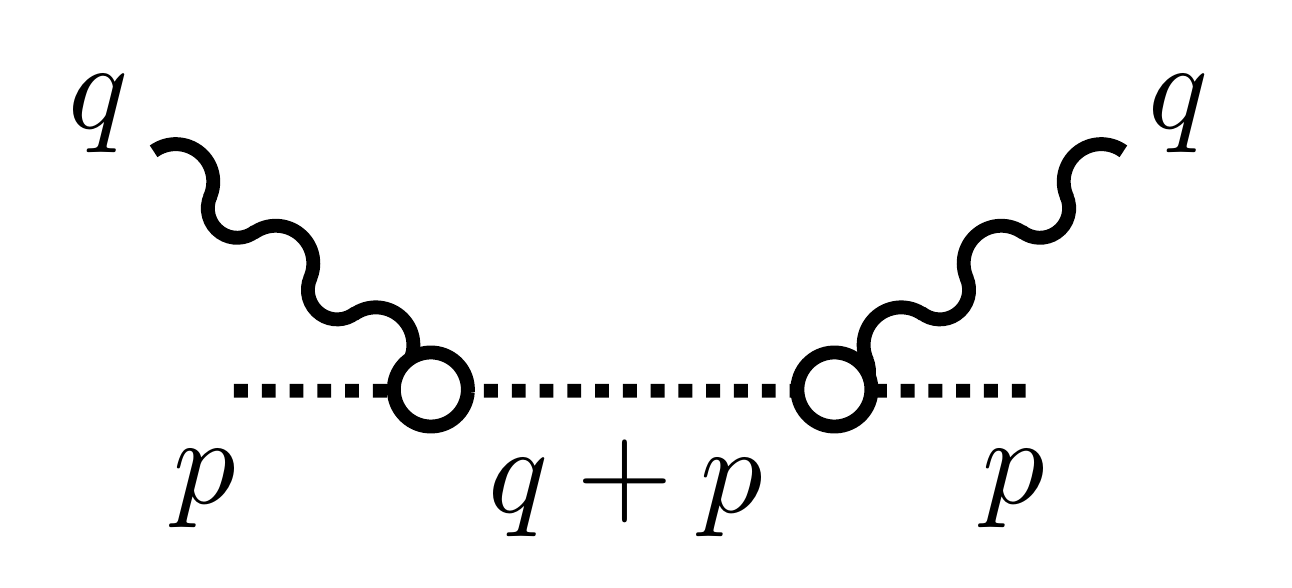}}
	\caption{(a) Feynman rules for the toy theory used in this work. The lighter particle, $\varphi$, is denoted by dashed lines, while the heavier particle, $\chi$, is denoted by the solid lines (b) The leading-order contribution to the matrix element ${\cal M}(\pmb \xi, \textbf p)$ with $\varphi$ external states.}
	\label{fig:rules_LO} 
\end{figure}

This EFT contains two scalar particles. One, $\varphi$, plays the role of the pion in QCD, and the other, $\chi$, is analogous to the nucleon or a heavy meson. These particles have pole masses $m_{\varphi}$ and $m_{\chi}$ respectively, satisfying $m_{\varphi} \ll m_{\chi}$. 
The states $\varphi$ and $\chi$ couple via a momentum-independent vertex defined diagrammatically in Fig.~\ref{fig:rules_LO}(a). We can then write the renormalized external current in terms of these fields,
\begin{equation}
\mathcal J(x) = \frac12  Z_{\varphi } g_{\varphi} \varphi^2 + \frac12 Z_{\chi } g_{\chi} \chi^2 + \frac12  Z_{\chi \varphi} g_{\chi \varphi} \chi^2 \varphi + \frac14  Z_{\chi \varphi\varphi} g_{\chi \varphi\varphi} \chi^2 \varphi^2 +\cdots \,.
\label{eq:currentdef}
\end{equation}
This generates the remaining Feynman rules shown in Fig.~\ref{fig:rules_LO}(a).

\begin{figure}[t]
	\centering
	\includegraphics[width =1\textwidth]{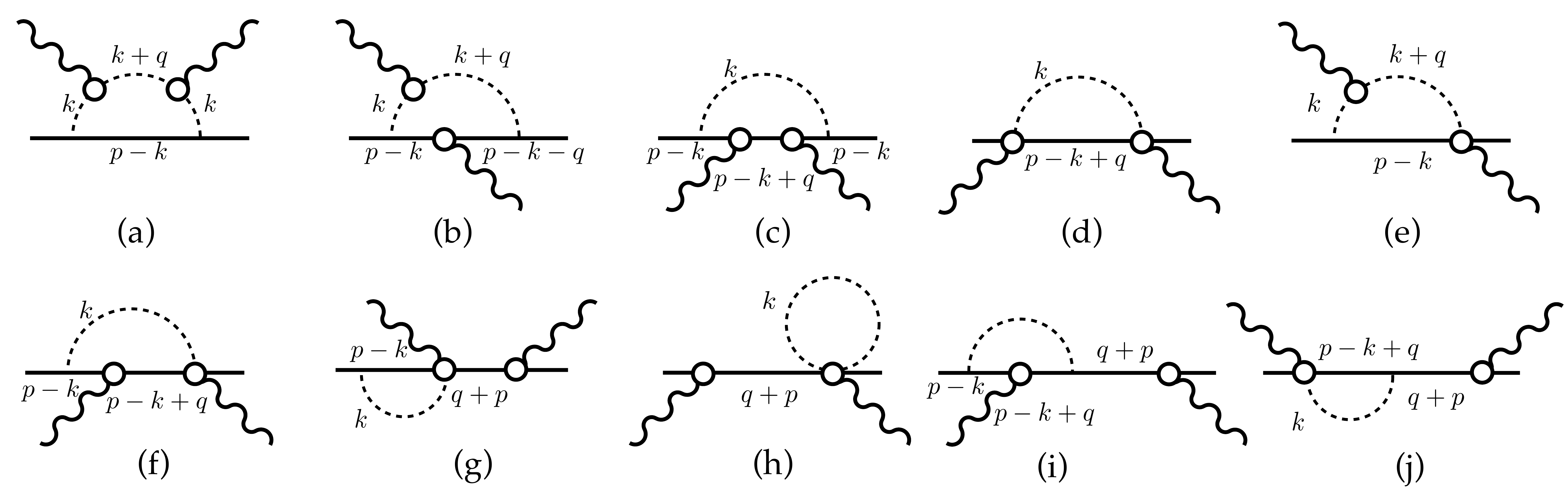}
	\caption{Contribution to the matrix element at next-to-leading-order when $\chi$ is the external state}
	\label{fig:chi_NLO} 
\end{figure}

Having defined the EFT, we can set up the general approach to determine finite-volume corrections contributing to the matrix elements of non-local operators. In this case the operator is composed of two identical currents, $\mathcal J(x)$, separated spatially by $\xi$. The matrix element is defined in the infinite-volume as: 
\begin{equation}
\mathcal M^{}_{\infty}(\pmb \xi, \textbf p)  \equiv \langle \textbf p \vert  \mathcal J(0, \pmb \xi) \mathcal J(0)    \vert \textbf p \rangle \,,
\label{eq:infME}
\end{equation}
where $\vert \textbf p \rangle$ is a single-particle state, either a $\varphi$ or a $\chi$, with momentum $\textbf p$. The contribution of any diagram, $d$, to $\mathcal M^{}_{\infty}(\pmb \xi, \textbf p)$ can be written in Euclidean space as,
\begin{equation}
\mathcal M^{(d)}_{\infty}(\pmb \xi, \textbf p) =  \int_{q_E} e^{i \textbf q \cdot \pmb \xi} \int_{k_{1,E}} \cdots \int_{k_{n-1,E}}   D^{(d)}_E(p_E,q_E, k_{1,E}, \cdots, k_{n,E}) \,,
\end{equation}
where $D^{(d)}_E(p_E,q_E, k_{1,E}, \cdots, k_{n,E})$ is the usual integrand that one would construct with Euclidean Feynman rules\footnote{Here we have introduced the notation $\int_{q_E} \equiv \int \frac{d^4 q_E }{(2 \pi)^4}$.}. Then, we can write a general expression for the finite-volume artifacts related to the matrix elements of spatially nonlocal currents. Using the Poisson summation formula, the finite-volume residue for diagram $d$ reads
\begin{align}
\delta \mathcal M^{(d)}_{L}(\pmb \xi, \textbf p) & \equiv \mathcal M^{(d)}_{L}(\pmb \xi, \textbf p) - \mathcal M^{(d)}_{\infty}(\pmb \xi, \textbf p)  \,, \\
&=   \sum_{ \textbf M \in \mathbb Z^{3n}\!/\{\textbf 0\} } \int_{K_E}  e^{i \textbf q \cdot \pmb \xi + i \textbf K \cdot L \textbf M}   D^{(d)}_E(p_E,K_{E}) \,, \label{eq:master}
\end{align}
where $\textbf M = \{ \textbf n, \textbf m_1, \cdots , \textbf m_{n-1} \}$ and the notation under the sum indicates that the only point omitted is when all three vectors vanish. Finally we have introduced $K_E = \{q_E, k_{1,E}, \cdots , k_{n-1,E} \}$.

\section{Results}

\begin{figure}[t]
	\begin{center}
		\includegraphics[width=\textwidth]{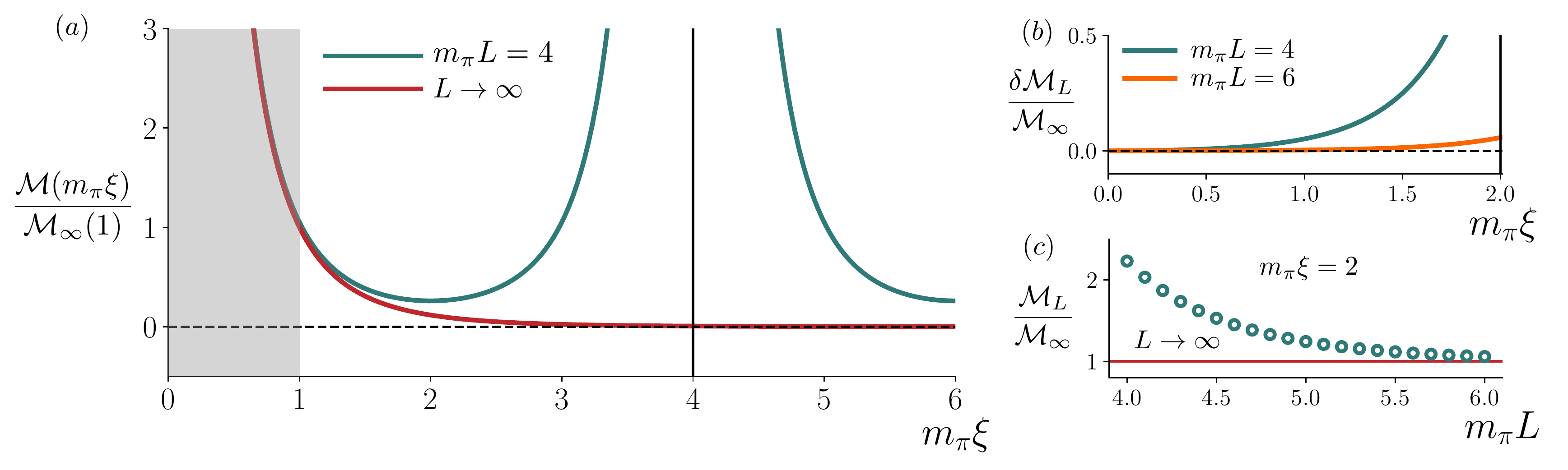}
		\caption{Infinite- and finite-volume behavior of matrix elements of spatially separated currents, from the tree-level result derived with pions as external states. Subfigure $(a)$ ilustrates the dependence of the matrix elements on the separation of the currents.  For $m_\pi\xi \lesssim 1$, indicated by the shaded region, high-energy scales are sample so that the effective field theory is expected to break down. Subfigure $(b)$ shows the relative difference between finite- and infinite-volume matrix elements, $\vert \mathcal M_L - \mathcal M_\infty \vert / \vert \mathcal M_\infty \vert$.	Finally, subfigure (c) shows the finite-volume matrix element, $\mathcal M_L$, as a function of $L$, together with its infinite-volume limit for a fixed separation $\xi$.}
		\label{fig:periodic}
	\end{center}
\end{figure}

\subsection{Light external states}

First, we focus on the case of matrix elements, ${\cal M}(\pmb \xi, \textbf p)$, with the lightest particle $\varphi$ as the external state. In this case the leading contribution to Eq.~\eqref{eq:infME} is given by the leading-order diagram, Fig.~\ref{fig:rules_LO}(b). Using 
Eq.~\eqref{eq:master}, we reach 
\begin{align}
\delta \mathcal M^{(\text{LO} )}_{L}(\pmb \xi, \textbf p) & =    g_\varphi^2\sum_{ \textbf n\neq 0 } \int_{q_E}  e^{i \textbf q \cdot (\pmb \xi + i L \textbf n)}   \frac{1}{(p_E+q_E)^2 + m_\varphi^2 } \,.
\end{align}

This integral can be solved analytically and written in terms of a modified Bessel function \cite{Briceno:2018lfj},
\begin{align}
\delta \mathcal M^{(\text{LO} )}_{L}(\pmb \xi, \textbf p) 
&  =   \frac{ m_{\varphi} g_\varphi^2}{4 \pi^2} 
e^{-i\textbf{p}\cdot  \pmb \xi }
\sum_{ \textbf n\neq 0}  
\frac{K_{1} \big (m_{\varphi} {|\pmb \xi+L\textbf n| } \big)}{|\pmb \xi+L\textbf n|} \,.
\label{eq:MLO_p}
\end{align}
From the asymptotic behavior of $K_{1}$, it is easy to see that the dominant finite-volume effect is given by the $\textbf n = - \hat{\pmb \xi}$ term. This term scales as
\begin{equation}\label{eq:m1bphi}
\delta \mathcal{M}^{(\text{LO})}_L ( \pmb \xi ,  \textbf p) =  \frac{ m_{\varphi} g_\varphi^2}{4 \pi^2}  e^{-i\textbf{p}\cdot  \pmb \xi }
\frac{K_{1} \big (m_{\varphi} {| L - \xi | } \big)}{|L -  \xi |}  \longrightarrow  \frac{ m_\varphi^2 g_\varphi^2}{4 \sqrt{2} \pi ^{3/2}} e^{-i\textbf{p}\cdot  \pmb \xi }  \, \frac{ e^{- m_\varphi (L- \xi)}}{ [m_\varphi (L- \xi) ]^{3/2}} \,,
\end{equation}
where the arrow indicates the asymptotic limit. As a result, we conclude that the overall volume scaling is by
\begin{equation}
{\delta \mathcal{M}^{(\text{LO})}_L ( \pmb \xi ,  \textbf p) }\propto    \frac{e^{- m_\varphi (L-  \xi) }}{(L- \xi)^{3/2}}  \,.
\label{eq:asymp_phi}
\end{equation} 

This is one of our key results, since it introduces a new scale parameterizing the size of finite-volume corrections, and it is related to the size of the operator. From Eq.~\eqref{eq:asymp_phi} we can read the infinite-volume prediction of this diagram by replacing $\vert L - \xi \vert$ with $\xi$.
In particular this implies that the diagram diverges in the limit $\vert \xi \vert \to 0$, as illustrated in Fig.~\ref{fig:periodic}. This is because our EFT is written in terms of hadrons and is only accurate for long distances $\xi>m_\varphi^{-1}$. Thus, we require $m_\varphi \xi  \gtrsim 1$, to guarantee that the finite- and infinite-volume matrix elements are accurately described by the EFT.

The plots in Fig.~\ref{fig:periodic} illustrate how to better understand Eq.~\eqref{eq:asymp_phi}. First, we note in Fig.~\ref{fig:periodic}(a) that the infinite-volume matrix element decays as a function of $\xi$, while its finite-volume counterpart is periodic and the difference between these two objects grows exponentially as $\xi$ approaches $L$. This behavior can be observed also in Fig.~\ref{fig:periodic}(b), where the finite-volume residue shows deviations of order $\sim10 \%$ for $\xi \sim L/4$ when $m_\pi L=4$. The plot in Fig.~\ref{fig:periodic}(c) complements these results, showing the finite-volume dependence of the matrix element on the size of the box for a fixed $\xi$. In this example, for a box of size of $m_\pi L=4$, there is a systematic uncertainty of $\sim 100 {\rm\, \%}$. However, these effects can be removed by performing a fit of the matrix elements to a decaying exponential in $L$ at fixed $\xi$.

\subsection{Heavy external states and general result}

Now, we focus on the case of matrix elements with the heavy particle, $\chi$, in the external states. The leading-order contribution in this case is given by diagram \ref{fig:rules_LO}(b), but exchanging the $\varphi$ particle by a $\chi$ particle. This gives, 
\begin{equation}\label{eq:m1bchi}
\delta \mathcal{M}^{(\text{LO})}_L ( \pmb \xi ,  \textbf p) =  \frac{ m_{\chi} g_\chi^2}{4 \pi^2} 
e^{-i\textbf{p}\cdot  \pmb \xi }
\frac{K_{1} \big (m_{\chi} {| L - \xi | } \big)}{|L -  \xi |}  \longrightarrow  
\frac{ m_\chi^2 g_\chi^2}{4 \sqrt{2} \pi ^{3/2}} e^{-i\textbf{p}\cdot  \pmb \xi }
\, \frac{ e^{- m_\chi (L- \xi)}}{ [m_\chi (L- \xi) ]^{3/2}} \,.
\end{equation}

This contribution is of order $\mathcal{O}\big(e^{- m_\chi (L- \xi)}\big)$ and is negligible in our power counting because $m_\chi \gg m_\varphi$ and $m_\varphi L \gg 1$.  Thus, in order to get the leading finite-volume corrections to a matrix element containing a $\chi$ external state we need to consider next-to-leading-order corrections coming from diagrams in Fig.~\ref{fig:chi_NLO}. The detailed calculation of these can be found in Ref.~\cite{Briceno:2018lfj}. We find that, for heavy external states, the finite-volume corrections scale as the standard exponential factor $ e^{- m_\varphi L  }$, but with a $\xi$ dependent pre-factor that can also enhance the corrections.

In general, the leading finite-volume effects for matrix elements of spatially separated currents can be written in a compact way
with two leading terms with the relevant scales encoding the finite-volume artifacts,
\begin{equation}
\delta \mathcal{M}_L  =  P_a(\pmb \xi, L)    e^{- M (L-  \xi) } +  P_b(\pmb \xi, L)    e^{- m_\pi L  }  + \cdots \,,
\label{eq:main_result}
\end{equation}
where $M$ represents the mass of the external state, and $P_a$ and $P_b$ are polynomial prefactors with terms scaling as $L^m/|L - \xi|^n$ and the ellipsis represents subleading exponentials. In the case of a pion external state, the first term scales as $e^{- m_\pi (L - \xi)}$ and is expected to dominate the volume effects as soon as $\xi$ becomes a non-negligible fraction of $L$. In the case of a heavy meson or nucleon, in which $M \gg m_\pi$, the second term is the dominant, assuming $\xi\ll L$ .

\section{Concluding remarks and impact in ongoing studies}

We have presented a study showing the first steps toward the understanding
of finite-volume artifacts in matrix elements of spatially non-local
operators. For this, we considered a toy EFT that involves 
two scalar particles, one light particle analogous to the pion and a heavier particle
analogous to a nucleon or heavy meson. The results of this work are summarized by Eq.~\eqref{eq:main_result}, where there are two terms that dominate the finite-volume artifacts. The first term in Eq.~\eqref{eq:main_result} introduces a new scale, $|L-  \xi|$, which encodes the finite-volume effects related to the size of the operator involved in the matrix element. The second term contains the usual scale dependence in the exponential, 
$e^{-m_\pi L}$, but has a potentially enhanced $\xi$ dependent pre-factor.
 	
There are several ongoing studies extracting PDFs from the lattice. These include investigations with two spatially separated currents \cite{Bali:2017gfr,Bali:2018spj} as well as studies using quark fields connected by a Wilson line \cite{Chen:2018xof,Chen:2018fwa,Lin:2018qky,Alexandrou:2018pbm,Alexandrou:2018eet, Orginos:2017kos}. The $m_\pi L$ values used in these studies vary over a wide range and, going forward, it will be important to check on a case by case basis to what extent finite-volume effects are relevant. In order to do so in a more systematic, quantitative way, one would need to use $\chi$PT (as suggested, for example, in Ref.~\cite{Bali:2018nde}). This will be more challenging for studies using a Wilson-line-based operator, where the mapping to a $\chi$PT description of the operator can be understood following Ref.~\cite{Kivel:2002ia}.

\acknowledgments

R.~A.~B.~and J.~V.~G.~acknowledge support from U.S. Department
of Energy Contract No. DE-AC05-06OR23177, under which Jefferson Science Associates,
LLC, manages and operates Jefferson Lab. J.~V.~G.~is supported in part by the U.S.~Department of Energy through DOE Contract No. DE-SC0008791 and also
through the JSA Graduate Fellowship Program.
C.~J.~M.~is supported in part
by the U.S.~Department of Energy through Grant No. DE-FG02-00ER41132.

\end{document}